# Defect-Free Carbon Nanotube Coils

*Nitzan Shadmi[†], Anna Kremen[‡], Yiftach Frenkel[‡], Zachary J. Lapin[§], Leonardo D. Machado[⊥], Sergio B. Legoas[∥], Ora Bitton[#], Katya Rechav[#], Ronit Popovitz-Biro[#], Douglas S. Galvão[⊥], Ado Jorio[∇], Lukas Novotny[§], Beena Kalisky[‡], and Ernesto Joselevich[†*]*

[†]Department of Materials and Interfaces and [#]Chemical Research Support, Weizmann Institute of Science, Rehovot, 76100, Israel. [‡]Department of Physics, Nano-magnetism Research Center, Institute of Nanotechnology and Advanced Materials, Bar-Ilan University, Ramat-Gan, 52900, Israel. [§]ETH Zürich, Photonics Laboratory, Zürich, 8093, Switzerland. [⊥]Instituto de Física "Gleb Wataghin", Universidade Estadual de Campinas, C. P. 6165, 13083-970 Campinas, Sao Paulo, Brazil. [∥]Departamento de Fisica, CCT, Universidade Federal de Roraima, 69304-000 Boa Vista, Roraima, Brazil. [∇]Departamento de Física, Universidade Federal de Minas Gerais, Belo Horizonte, MG, 31270-901, Brazil.

ABSTRACT

Carbon nanotubes are promising building blocks for various nanoelectronic components. A highly desirable geometry for such applications is a coil. However, coiled nanotube structures reported so far were inherently defective or had no free ends accessible for contacting. Here we demonstrate the spontaneous self-coiling of single-wall carbon nanotubes into defect-free coils of up to more than 70 turns with identical diameter and chirality, and free ends. We characterize the structure, formation mechanism and electrical properties of these coils by different microscopies, molecular dynamics simulations, Raman spectroscopy, and electrical and magnetic measurements. The coils are highly conductive, as expected for defect-free carbon nanotubes, but adjacent nanotube segments in the coil are more highly coupled than in regular bundles of single-wall carbon nanotubes, owing to their perfect crystal momentum matching, which enables tunneling between the turns. Although this behavior does not yet enable the performance of these nanotube coils as inductive devices, it does point a clear path for their realization. Hence, this study represents a major step toward the production of many different nanotube coil devices, including inductors, electromagnets, transformers and dynamos.

Extensive research has been devoted to exploring the potential of carbon nanotubes for the assembly of various nanoelectronic components, including transistors,[1] diodes,[2] resistors,[3] capacitors[4] and interconnects.[5] One important component yet to be demonstrated is a coil. Carbon nanotube coiled structures reported so far were based on periodic defects, which induce curvature,[6-7] but also scattering and high resistance,[8-9] or had no free ends available for electrical contacting.[10-11] In order to produce carbon nanotube coils that are suitable for electronic applications, we investigate the coiling of defect-free carbon nanotubes and the properties of the resulting nanotube coils.

Coiling is a general phenomenon that can be observed at very different scales in falling flexible rods,[12] such as cables, ropes and spaghetti, as well as in viscous jets,[13] such as when pouring honey or shampoo. In these macroscopic cases, coiling is usually driven by gravity, while self-affinity plays a secondary role. In microscopic systems, coiling is often driven by self-affinity, for instance by the addition of a condensing agent such as spermidine to DNA plasmids[14], or by dipolar interactions in ZnO nanobelts, leading to their epitaxial self-coiling into single-crystal rings[15]. In the case of carbon nanotubes, coiled structures were previously obtained by the formation of periodic structural defects in the gas phase[6], or by the aggregation of many nanotubes into toroidal bundles.[10-11, 16] Incidental observation of coiling at the end of individual single-wall carbon nanotubes was also reported[17], and believed to take place in free space prior to landing on a substrate, but no focused studies on this phenomenon have been reported. Overall, none of these coiled structures were suitable for electronic applications, either because of their large defect-induced resistance or because their ends were not well separated from the bundle to facilitate their selective connection to electrodes.

Here we report for the first time the self-coiling of single-wall carbon nanotubes into defect-free coils with fully accessible ends. This configuration allowed us to connect the two ends of each coil to electrodes, and thus to characterize its electrical properties (Fig. 1a,b). As opposed to previous coiled structures, where coiling takes place in the gas phase or in liquid suspension, our self-coiling process takes place on regular silicon substrates, making the coils ready for integration into a series of potential devices. Another unique feature of these structures is that all the parallel nanotube segments in the coil have the same diameter and chirality, and are thus hexagonally packed as a single-crystal. Long-sought single-crystals of single-wall carbon nanotubes were initially reported, but later found to be artifact[18]. Our optical and electronic characterization of true single-crystals of single-wall carbon nanotubes reveals a higher coupling than in regular ropes of single-wall carbon nanotubes with different chiralities.

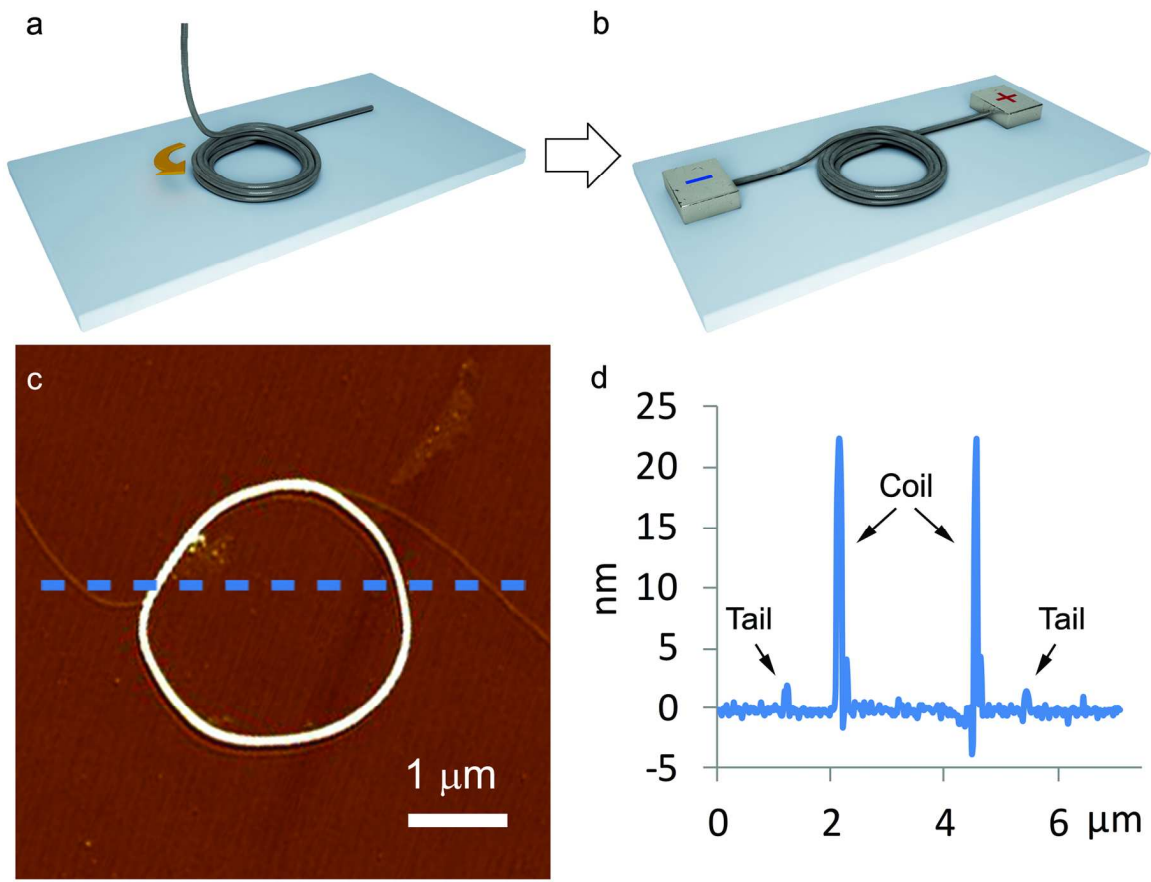

**Figure 1.** Nanotube coils: concept and production. **a**, Schematic representation of the formation of a single-wall carbon nanotube coil. **b**, Schematic representation of the formed coil with its two free ends connected to electric leads. **c**, AFM height image of a nanotube coil. The blue dashed line shows the position of the topographic cross-section shown in **d**. **d**, The heights of the two sides of the coil are 22.4±0.3 nm (left) and 22.2±0.3 nm (right), and the height of the free ends of the coil is 1.6±0.3 nm.

The single-wall carbon nanotubes were grown on Si/SiO$_2$ substrates from thin stripes of Fe nanoparticles patterned on a supporting layer of SiO$_2$, known to promote suspended growth of long (>100 μm) single-wall carbon nanotubes[19]. Initially, a relatively low flow rate of 70-500 sccm was used, to offer minimal perturbation of the vertical fall, although a similar yield was attained when using flows of up to 2000 sccm (see further discussion in Supporting Information). This procedure produced tens to over a hundred nanotube coils per sample, which can be efficiently mapped by scanning electron microscopy (SEM, Fig. S1). The coils can be distinguished from simple loops by the entry and exit points of the nanotubes around the closed ring part. If these points are different (Fig. S1), then the structure is a coil with more than one turn. The diameter of the coils is typically 2-4 μm, although it can be as large as 10 μm, and as small as 1 μm. The coil often has a higher SEM contrast than the individual nanotube (Figs. 2a,d, S1). Atomic force microscopy (AFM) is used to verify the topographic height of each selected coil (Figs. 1c, 2f,i) relative to that of its free ends, which provides a rough estimation of the number of turns. Absolute determination of the number of turns in the coil is done by cross-sectional transmission electron microscopy (TEM). For this, we cut a thin (50-100 nm) lamella across the coil using a focused-ion beam (FIB), and observe the two cross-sections at opposites regions of the coil under the TEM (see Methods section). Each cross-section shows a densely packed hexagonal lattice of identical nanotubes, each corresponding to one turn of the coil (Figs. 2b,c,e,g,h,j, 4e and S2).

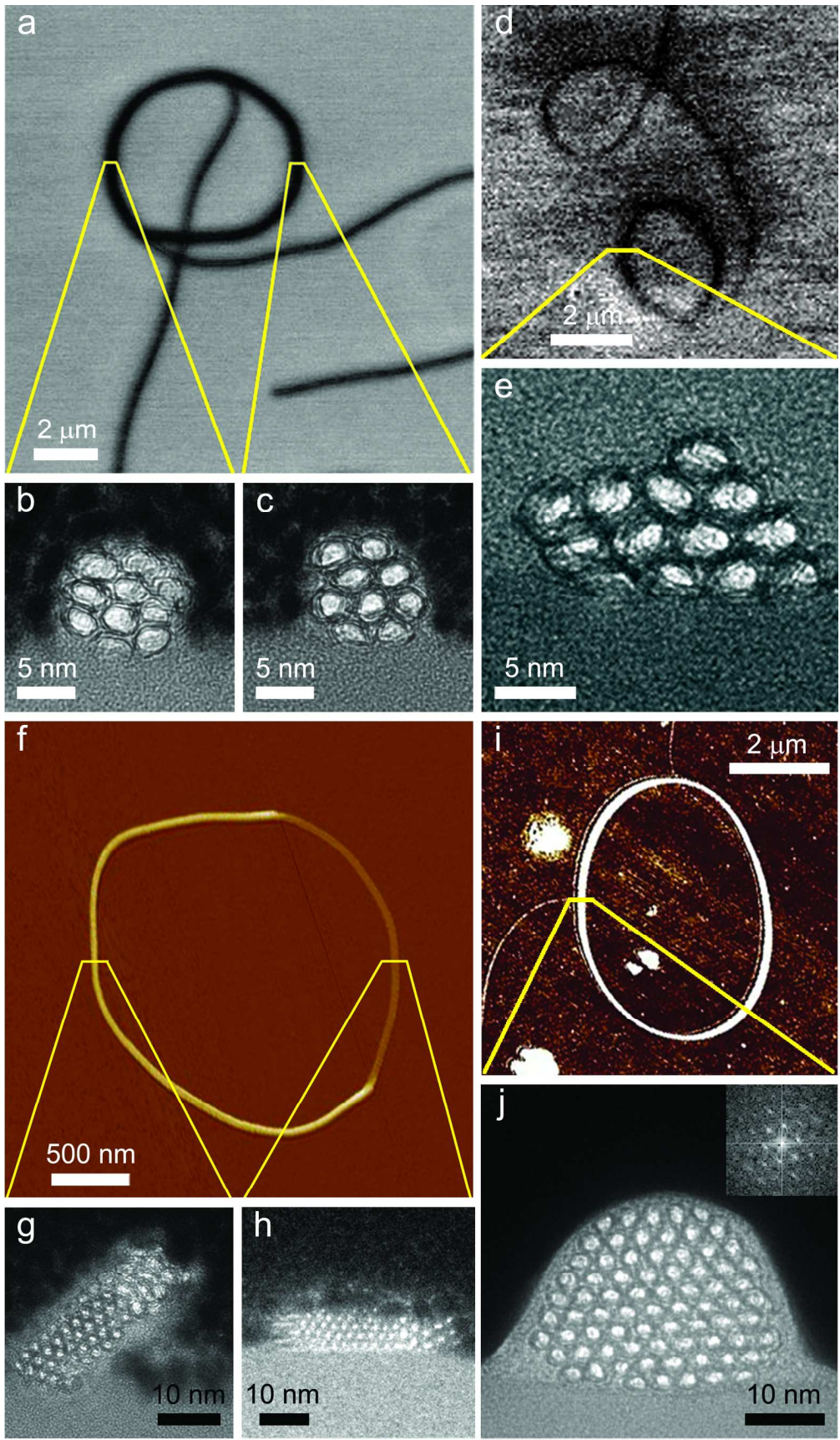

**Figure 2.** Structural characterization of the nanotube coils. **a**, **d**, **f** and **i**, SEM images (**a** and **d**) and AFM height images (**f** and **i**) of coils from which a lamella was cut. **b**, **c**, **e**, **g**, **h** and **j**, TEM images of the cross-section. Yellow lines show the location of the matching cross-section. By counting the turns in the corresponding TEM images (**b**, **c**, **e**, **g**, **h**, and **j**), we determine that the coils shown in **a**, **d**, **f** and **i** have 9.5, 13, 57 and 74 turns, respectively. The inset in **j** shows the fast Fourier transform of the cross-section, which gives a lattice parameter of 2.6±0.2 nm. Accounting for a van der Waals distance of 0.34 nm, this gives a diameter of 2.3±0.2 nm, matching a single-wall carbon nanotube.

We imaged cross-sections of nine coils, and accurately counted their number of turns, which ranged from 4 to 74 (Figs. 2b,c,e,g,h,j, 4e and S2). Several coils had rectangular-shaped cross-sections (e.g. Fig. 2g-h), similar to a nanoribbon, suggesting that there is a preference for the turns to form on one specific side of the coil. In this case, the state of the coil can be mixed, with part of it lying on its wide side, and another part on its narrow side (Fig. 2f-h). The cross-sectional TEM images of these parts are consistent with their AFM topography, and the SEM images also show alternate left and right twists between these parts of the coil (Fig. 3 and see further discussion in Supporting Information).

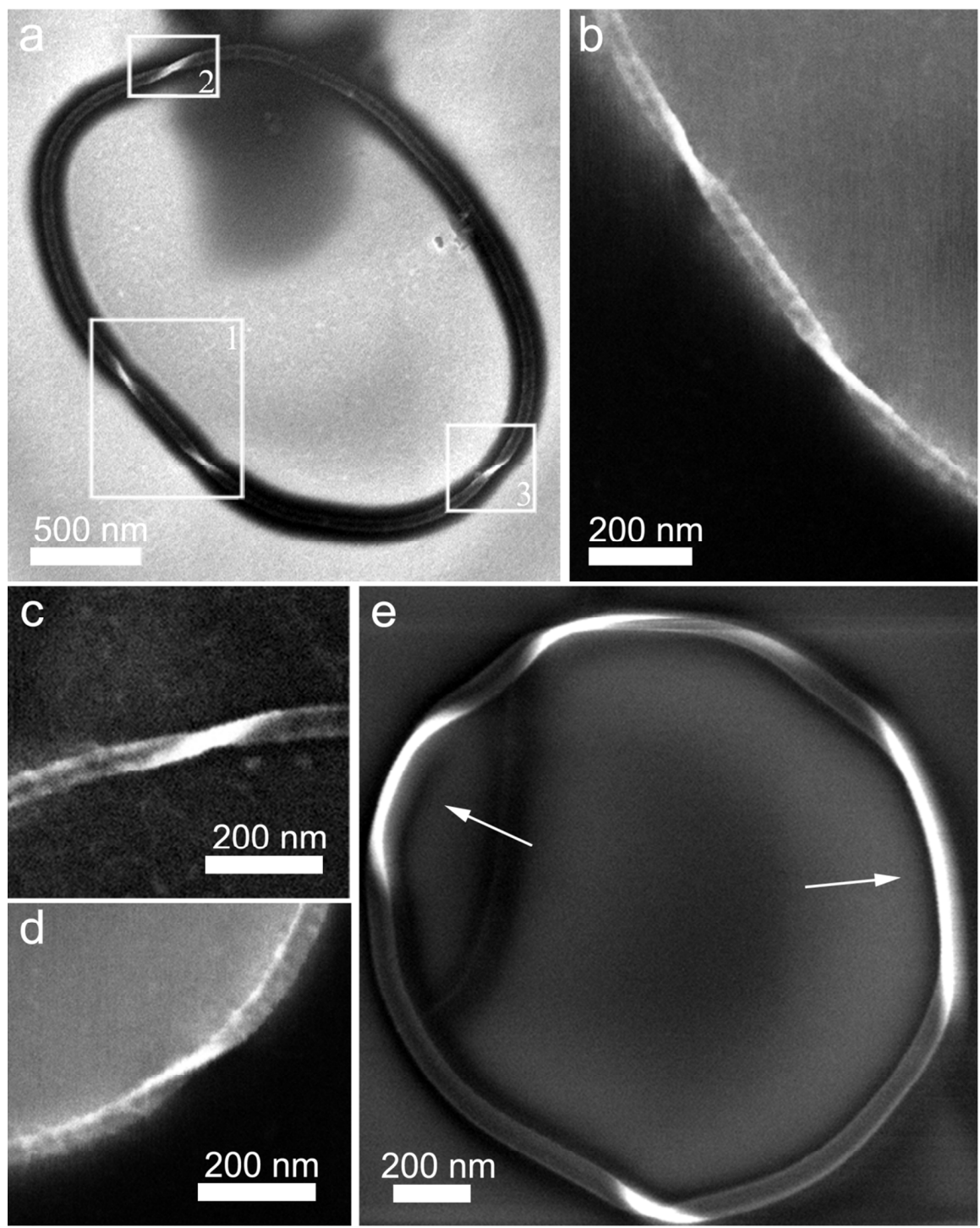

**Figure 3.** Occurrence of twists in ribbon-like carbon nanotube coils. **a**, SEM image of a coil with twists. **b-d**, Zoomed-in images of the areas of the twists, marked by white frames 1-3,

respectively. **e**, A coil with some vertically aligned sections (marked by white arrows). The horizontally aligned sections are clearly broader than the vertically aligned ones.

The formation of defect-free carbon nanotube coils can be qualitatively rationalized in the framework of a "falling spaghetti" mechanism, initially proposed for the self-organization of nanotube serpentines[19-21] and more recently extended to other curved structures[22]. In this mechanism, the nanotube first grows up from the substrate and later falls making wiggles, which can be directed both by the substrate and the gas flow. The fall is driven by van der Waals interactions with the substrate while subject to dynamic instabilities. Depending on the relative velocity of the forward and downward components of the nanotube motion, four different structures can be expected, including flow-aligned, serpentine, looped and coiled geometries (Fig. 4a-e). A quite similar macroscopic mechanism was recently proposed and tested for the coiling of elastic rods on rigid substrates[12]. In the case of a carbon nanotube, its van der Waals interaction with itself is particularly strong[23], and predominance of the downward motion increases the probability of coiling of the nanotube around itself (Fig. 4b, and see further details in Supporting Information).

More quantitative insight into the mechanism of self-coiling of single-wall carbon nanotubes on the amorphous Si/SiO$_2$ was obtained from atomistic molecular dynamics simulations (see Supporting Information for detailed description and discussion), similar to those recently performed to model the formation of nanotube serpentines on crystalline substrates[20]. Our simulations fully reproduce the spontaneous formation of carbon nanotube coils, and describe the evolution of strain, van der Waals energy and kinetic energy during the self-coiling process (Fig. 4f-i, and Movies S1 and S2). These simulations indicate that the van der Waals interaction is much stronger than the strain energy, so the overall energy of the coil is lower than that of a straight nanotube. Hence, once the first turn is formed, the self-coiling proceeds steadily, until disturbed by a sufficiently strong fluctuation.

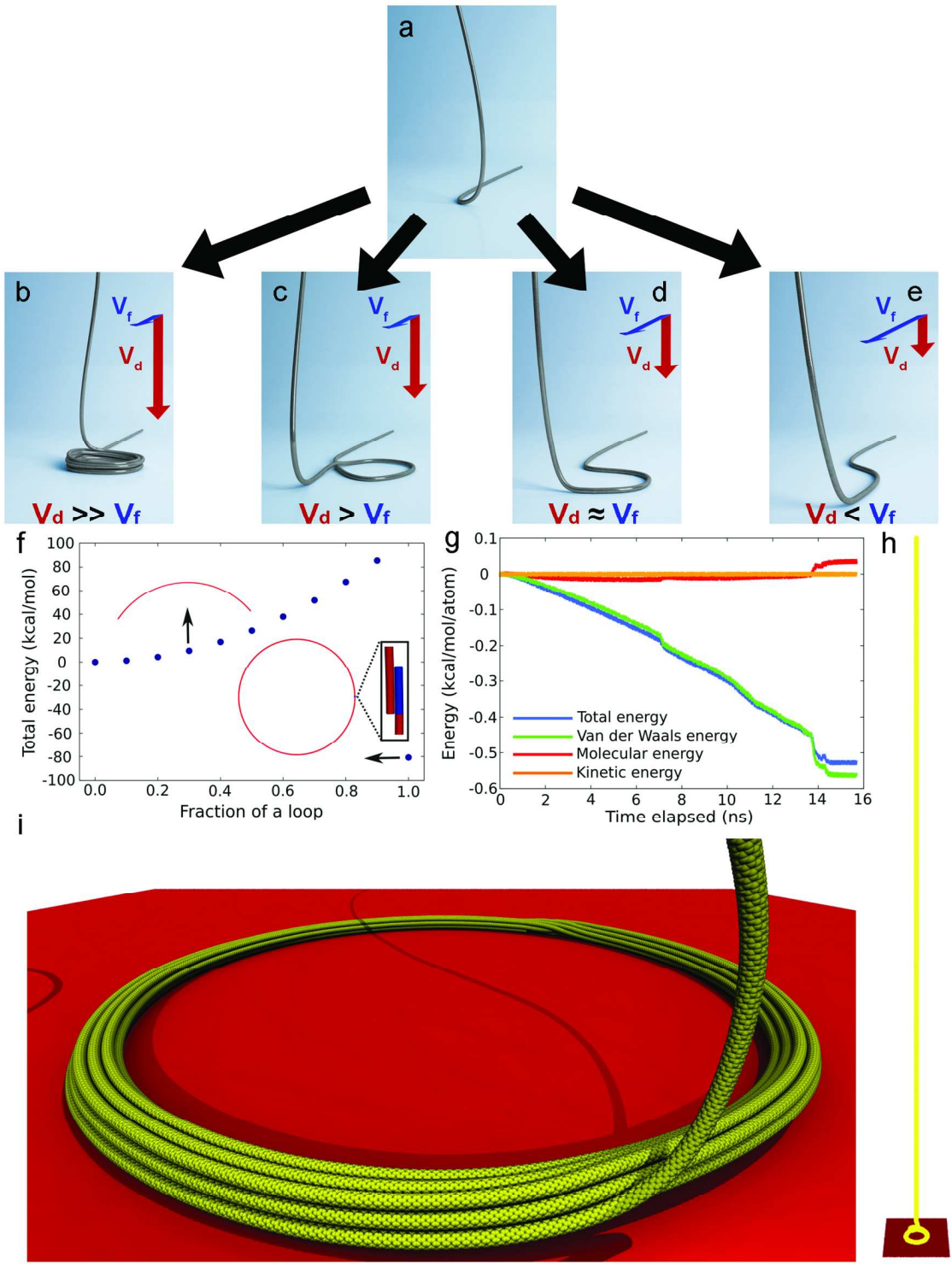

**Figure 4. Formation mechanism of defect-free carbon nanotube coils. a**, A nanotube has grown above the surface, and begins to deposit onto it, creating a suspended half loop. **b-e**, Four possible geometries that may form, depending on the relative forward (in the direction of the gas flow) and downward (in the direction of the substrate) velocities, $v_f$ and $v_d$, respectively. **b**, $v_f$ is negligible in comparison with $v_d$: the nanotube coils like a falling rope, forming multiple turns. **c**, $v_f$ is sufficiently lower than $v_d$: the nanotube forms a single loop. **d**, $v_f$ is roughly the same as $v_d$: the suspended half loop falls in the gas flow direction, and a serpentine U-turn is formed. **e**, $v_f$ is higher than $v_d$: the nanotube aligns in the direction of the gas flow, forming a straight segment. **f**, Energy needed to bend a 3 μm (13,0) nanotube to a fraction of a turn. As curvature increases, so does the elastic energy cost. When a full turn is completed, there is a small 7 nm nanotube overlap length, shown zoomed in at the inset. Because of the van der Waals attraction at the overlapped region, the total energy of the full turn is smaller than that of the straight tube – that was defined as zero. **g**, Evolution of the total, van der Waals, molecular and kinetic energies of a 3 μm (6,0) nanotube as it assembles into a coil. Initial energies were defined as zero. **h**, Far view of the initial structure used in the simulation, with nanotube diameter greatly exaggerated to improve visualization. **i**, The structure at t=12.2 ns. See also Movies S1 and S2.

The structural perfection of our carbon nanotube coils was evaluated by Raman spectroscopy (Figs. 5a and S3 as well as Supporting Information). The narrow Raman lines (FWHM below 13 cm$^{-1}$)[24] and the negligible intensity of the disorder-induced D-band (~1350 cm$^{-1}$, average $I_D/I_G$ = 0.04±0.04 for 9 coils)[25], are both indicative of a very low density of structural defects. Moreover, imaging of the G-band shows that the coil and its free ends are all simultaneously in resonance with the same laser energy, indicating that all the nanotube maintains a constant diameter and chirality along the entire coil.

An ideal electromagnetic coil is made up of a coil along which an electric current is passed, and a uniform magnetic field is generated, with an intensity that is proportional to the number of turns the current passes. To assess the functionality of our defect-free carbon nanotube coils for the construction of inductive devices, we determine the effective number of turns that a current passes before it shorts by tunneling between adjacent turns. This was done first by measuring the current as a function of bias and gate voltage, and then by measuring the magnetic field over the coil while applying a voltage between the free ends of the coil.

On each selected coil (Fig. 5b) we performed two four-point probe electrical measurements, first on one free end of the coil (Fig. 5c), and second between the two ends of the coil (Fig. 5d). From the first measurement while gating, we found these nanotubes to be p-type semiconducting, and also determined the nanotube length-resistivity (50-100 k$\Omega$/µm). From the second measurement and this length-resistivity, we calculated the effective length of coiled nanotube followed by the current. Taking that length, and dividing it by the circumference of the coil, we roughly estimate the effective number of turns that the current passed through to be 0.4±0.3 (i.e. half a turn) for the four coils tested. This suggests that the electrical behavior of the coil is dominated by tunneling between its turns ("inter-turn tunneling").

In order to determine more directly the effective number of turns the current flows through, we measured the magnetic field over the biased coil by scanning superconducting quantum interference device (SQUID) microscopy correlated with AFM (Fig. 5f,k,p). We compare the measured magnetic field generated by the current flow at the centre of the coil to the field generated at uncoiled parts of the nanotube and the electrodes. The field at the coil centre should appear as enhanced or revered to the overall field, depending on the coiling direction of the nanotube coil. We reconstructed the current path from magnetic flux measurement of the SQUID (Fig. 5i,n,s) and compared them to the simulations. Based on the SEM and AFM images of the coils (Fig. 5f,k,p) we resolved the direction of coiling for each coil and calculated the expected magnetic flux image by the Biot-Savart law (Fig. 5i,j,n,o,s,t). Three different coil samples were prepared, all of which were p-type semiconducting. The magnetic flux at the centre of the coils was measured under various bias and gate voltages. By comparing the results (Fig. 5g,h,l,m,q,r) with the simulations, we determined the effective number of turns to be less than one, consistent with the result obtained from the resistance measurements (Fig. 5c-d).

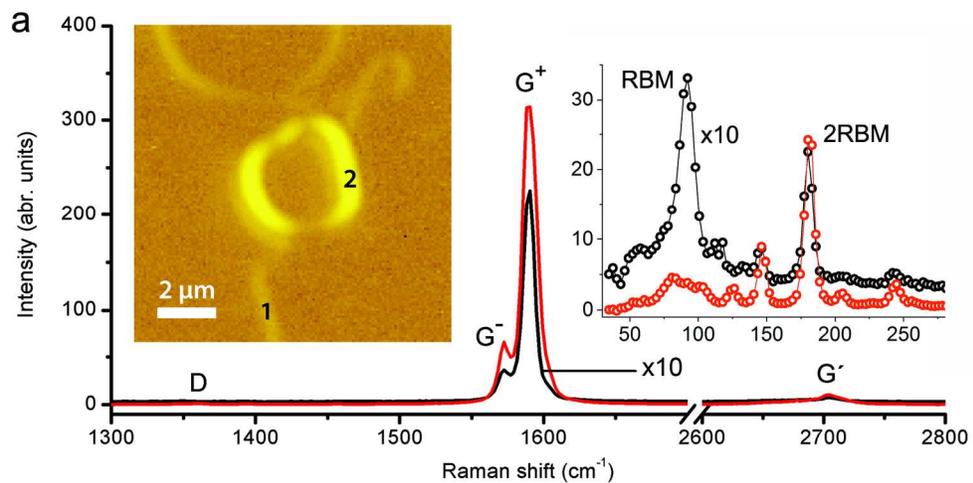
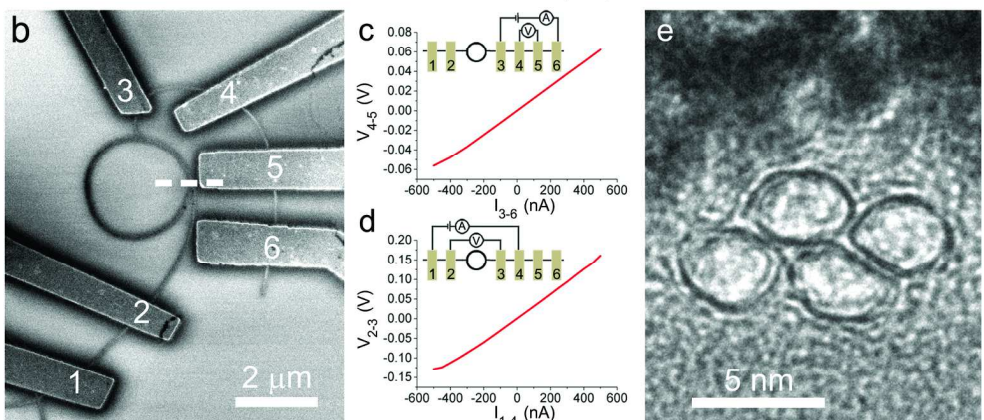
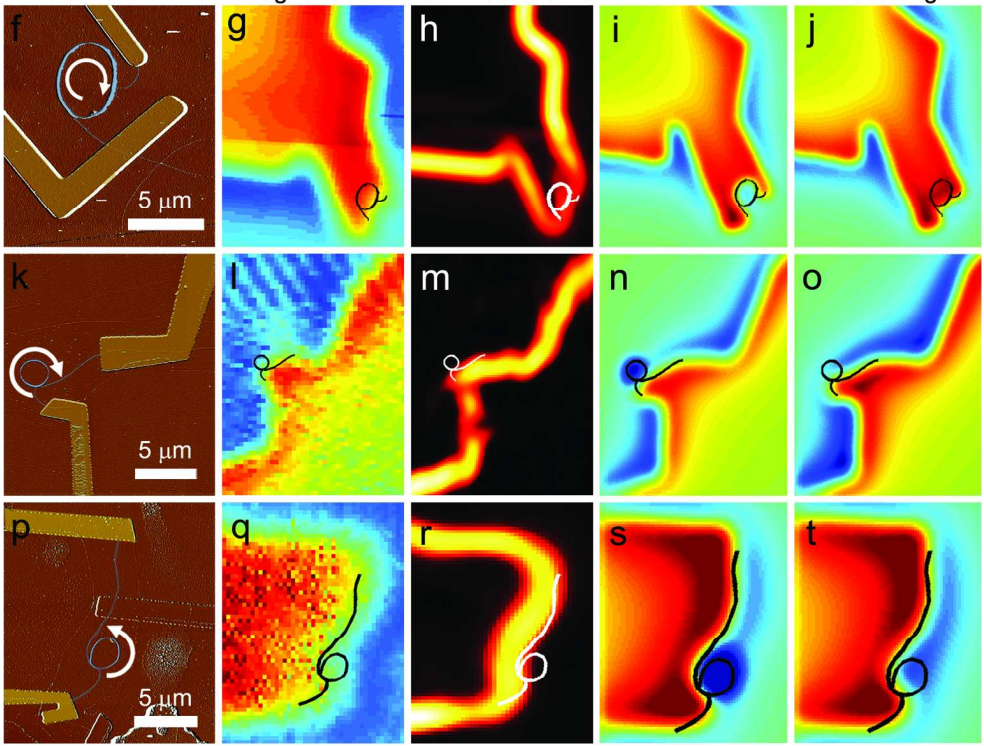

**Figure 5. Optical, electrical and magnetic characterization. a**, Overlaid Raman measurements of the free end (black spectrum, position 1 in left inset) and coil (red spectrum, position 2 in left inset) segments of the same nanotube (identified as semiconducting[26]). Left inset shows a G-band Raman image of the coil, where the coil section, which is composed of the signal of several nanotube turns, gives a significantly stronger signal than the free end. There are no apparent bundling induced changes in high (> 1000 cm$^{-1}$) modes. Right inset is the low frequency region of the spectrum, showing the radial breathing mode (RBM) peak from the single-wall carbon nanotube at ~90 cm$^{-1}$, obtained from the free end segment, as well as another peak at ~180 cm$^{-1}$, which is probably an RBM overtone.[27] Interestingly, the RBM peak is highly broadened in the coil segment, a result that has been predicted theoretically as due to van der Waals interactions[28], but never measured in carbon nanotube bundles formed by different (n,m) tubes. **b**, SEM image of a measured nanotube with Pd electrodes. **c** and **d**, Four-point probe measurements on the free end and coil segments of the nanotube, respectively. These measurements were performed using a gating voltage of -10 V, as the nanotube was found to be p-type. **e**, Cross-sectional TEM of the lamella taken at the position marked by a white dashed line in **b**. The image shows that the coil comprised 3 complete turns, and an additional ~1/4 turn. **f**, **k** and **p,** AFM images of coils contacted by Pd electrodes. The nanotubes and the source and drain electrodes are falsely colored for emphasis. The boundary lines visible in **k** and **p** surrounding the area of the nanotubes are due to the oxygen plasma treatment used to remove other nanotubes on the sample (see methods section for more details). The white arrows indicate the coiling direction of the nanotubes. **g**, **l** and **q,** The magnetic flux captured by scanning SQUID microscopy, at 4K, AC current of 100 nA, 100 nA and 20 nA, and gate voltage of -10 V, -10 V and -7 V, respectively. Color bar spans 0.6, 0.6 and 0.3 m$\Phi_0$, respectively. Red (blue) represents positive (negative) flux

response. **h**, **m** and **r**, The respective current paths reconstructed from the flux data (a.u.). **i**, **n** and **s**, The respective magnetic flux images calculated for current flow from electrode to electrode passing through one effective turn of the nanotube coil. **j**, **o** and **t**, The respective magnetic flux images calculated for the shortest current path. The outline of the measured nanotube in black or white was added to **g-j**, **l-o** and **q-t** for reference.

Our results demonstrated that self-coiling of single-wall carbon nanotubes leads to formation of defect-free coils, which are highly conductive and have free ends available for contacting, as required for the assembly of functional devices. However, the electrical behavior of the defect-free nanotube coil is dominated by inter-turn tunneling rather than end-to-end conduction. This finding may seem surprising considering that intertube tunneling in single-wall carbon nanotube ropes is usually weaker than end-to-end conductance[29]. One way of explaining this behavior is that in a regular bundle of single-wall carbon nanotubes, each nanotube has a different chirality, and hence the intertube tunneling involves a large change in crystal momentum, which cannot be acquired by coupling to phonons under low bias conditions[30]. However, in our defect-free coils, all the parallel tubes, each being a different turn of the same nanotube, have exactly the same diameter and chirality. Therefore, the electrons can easily tunnel from one turn to another without changing their crystal momentum[31]. Consequently, the nanotube shorts with itself as an unsheathed wire. On one hand, this may be a disappointing conclusion from the perspective of the envisaged inductive applications. On the other hand, it is a mandatory lesson to be learned, which can lead to progress when knowledgeably addressed. How can one effectively sheathe single-wall carbon nanotubes to inhibit intertube shorting? One possible solution that we propose is using double-wall carbon nanotubes[32], so that the inner tube is sheathed by the outer wall having a different chirality. This will require and inspire many new and intriguing theoretical and experimental studies. In any case, we believe that the formation and characterization of defect-free carbon nanotube coils demonstrated here represents a major development toward the production of a large variety of nanotube coil devices.

ASSOCIATED CONTENT

**Supporting Information**

(1) Additional SEM, AFM and TEM images of carbon nanotube coils and their cross-sections. (2) Raman spectra of defect-free nanotube coils and their analysis. (3) Discussion of the effect of gas flow on the formation of defect-free coils. (4) Discussion on the orientation of ribbon-like nanotube coils and the occurrence of twists. (5) Movies of molecular dynamics simulations of the formation of defect-free carbon nanotube coils, and their analysis. (6) Electrical characterization of the effects of bundling in nanotube coils. (7) Details of the experimental methods. This material is available free of charge via the Internet at http://pubs.acs.org.

AUTHOR INFORMATION

**Corresponding Author**

* E-mail: ernesto.joselevich@weizmann.ac.il. Phone: +972-8-9342350. Fax: +972-8-9344138.

ACKNOWLEDGMENT

We thank John R. Kirtley and Kathryn A. Moler for their help with the scanning SQUID measurements at Stanford University, Amos Sharoni and Tony Yamin for their help with Nb sputtering, Palle von Huth and David Tsivion for assistance with the FIB, Avishai Benyamini, Jonah Waissman and Assaf Hamo for assistance with low-temperature electrical measurements, and David Rakhmilevich and Shahal Ilani for helpful discussions. This research was supported by the Israel Science Foundation, the Helen and Martin Kimmel Center for Nanoscale Science, the Moscowitz Center for Nano and Bio-Nano Imaging, and the Perlman Family Foundation. E.J. holds the Drake Family Professorial Chair of Nanotechnology. B.K. acknowledges support of the European Research Council ERC-2014-STG- 639792, ISF (ISF #1102/13), and CIG FP7-PEOPLE-2012-CIG-333799. L.N. acknowledges financial support by the Swiss National Science Foundation (grant 200021_149433). L.D.M. and D.S.G. thank CNPq and CCES for financial support through FAPESP/CEPID Grant # 2013/08293-7. The Scanning SQUID Microscope was constructed with support from the National Science Foundation DMR-0957616 and is part of the Stanford Nano Shared Facilities.